%% file: conference_101719.tex
\documentclass[conference]{IEEEtran}
\IEEEoverridecommandlockouts
\usepackage{cite}
\usepackage{amsmath, amsthm, amssymb}
\usepackage{multirow}
\usepackage{algorithmic}
\usepackage{graphicx}
\usepackage{textcomp}
\usepackage{xcolor}
\usepackage{tikz}
\usetikzlibrary{intersections}
\def\BibTeX{{\rm B\kern-.05em{\sc i\kern-.025em b}\kern-.08em
    T\kern-.1667em\lower.7ex\hbox{E}\kern-.125emX}}
\usepackage{url}
\usepackage[left=1.62cm,right=1.62cm,top=1.9cm,lmargin=2.7cm,rmargin=2.7cm]{geometry}
\begin{document}

\title{Decentralized Exchanges: The Profitability Frontier of Constant Product Market Makers}

\author{\IEEEauthorblockN{Tobias Bitterli}
\IEEEauthorblockA{\textit{Center for Innovative Finance} \\
\textit{University of Basel}\\
Basel, Switzerland \\
tobias.bitterli@unibas.ch}
\and
\IEEEauthorblockN{Fabian Schär}
\IEEEauthorblockA{\textit{Center for Innovative Finance} \\
\textit{University of Basel}\\
Basel, Switzerland \\
f.schaer@unibas.ch}}

\IEEEoverridecommandlockouts
\IEEEpubid{\makebox[\columnwidth]{Forthcoming in IEEE CryptoEx 2023. Please cite published version. \hfill} \hspace{\columnsep}\makebox[\columnwidth]{ }}

\maketitle

\IEEEpubidadjcol

\begin{abstract}
In this paper we analyze constant product market makers (CPMMs). We formalize the liquidity providers' profitability conditions and introduce a concept we call the \emph{profitability frontier} in \textit{xyk}-space. We study the effect of \emph{mint} and \emph{burn} fees on the profitability frontier, consider various pool types, and compile a large data set from all Uniswap V2 transactions. We use this data to further study our theoretical framework and the profitability conditions. 
We show how the profitability of liquidity provision is severely affected by the costs of \emph{mint} and \emph{burn} calls relative to the portfolio size and the characteristics of the trading pair.
\end{abstract}

\begin{IEEEkeywords}
Automated Market Maker, Blockchain, Constant Product Market Maker, Decentralized Exchange, Decentralized Finance, Uniswap.
\end{IEEEkeywords}

\section{Introduction}

Constant product market makers (CPMM) are smart contract-based liquidity pools that contain two distinct assets. They serve as neutral exchange infrastructure and employ an endogenous pricing model, based on the proportion of their token reserves. Generally speaking, the greater a CPMM's reserve of token $x$ in relation to the CPMM's reserve of token $y$, the higher the relative price of token $y$. 

Anyone can become a liquidity provider (LP) by contributing tokens to the pool contract in line with the current pool ratio. They can later redeem their share and close their position by withdrawing their proportional stake of the pool's liquidity.
LPs essentially act as passive market makers. Their allocation (i.e., ratio between $x$ and $y$ token holdings) changes whenever anyone performs a swap using the CPMM. To compensate LPs for their opportunity costs and the risks of passive market making, they earn a portion of the trading fees. 

In this paper we take a closer look at these effects and analyze the profitability of CPMM LPs. In Sections \ref{sec:relatedWork} and \ref{sec:framework} we lay the foundation by summarizing related literature and introducing the formal framework for our analysis. In Section \ref{sec:profitabilityAnalysis} we propose a concept we call the \emph{profitability frontier} in the $xyk$-space and study the effects of \emph{mint} and \emph{burn} fees. Moreover, we consider the effects of various pool types. In Section \ref{sec:data} we provide empirical evidence from Uniswap V2 pools and show how relative fee size and pool type affect the profitability conditions. In Section \ref{sec:conclusion} we discuss our results and conclude.

\section{Related Work}
\label{sec:relatedWork}


To the best of the authors' knowledge, the concept of a blockchain-based automated market maker (AMM) was first proposed by \cite{Lu2017} and \cite{hbbBancor2018}. \cite{adamsUniswap2018} introduced a simplified model based on the work of \cite{buterin2017}, which was later formalized by \cite{zcpFormal2018} and extended by \cite{uniswapV2_2020}. \cite{martinelliBalancer2019} generalized the concept to allow for various token weights and pools with more than two assets. Furthermore, modifications for specific use-cases, such as stablecoin exchanges, have been proposed by \cite{egorov2019stableswap}. Additionally, \cite{uniswapV3_2021} introduced the concept of concentrated liquidity. 

There is a significant body of literature regarding the design of AMMs, as demonstrated by \cite{schaer21, Mohan2020}, as well as their properties, outlined in \cite{Xu_2022, bartoletti2021,  angerisUniswap2019}. \cite{angerisCFMM2020} propose a general framework for the analysis of CPMMs as a major subset of AMMs. 
Numerous articles discuss the efficiency of AMM designs with respect to price discovery \cite{Angeris_12_2020, pourpouneh2020automated} and compare price efficiency to centralized exchanges \cite{barbon2020, leharParlour2022}. \cite{leharParlour2022} further characterize equilibrium liquidity pools and find no long-lived arbitrage opportunities. \cite{angerisUniswap2019} conduct agent-based simulations to demonstrate that AMMs can theoretically be used as sound price oracles. \cite{evans2020liquidity} uses time-varying and stochastic weights to replicate the payoffs of financial derivatives. \cite{capponiJia2021} investigate the market microstructure of AMMs. 

Many papers study the profitability of providing liquidity to AMMs. For instance, \cite{aoyagi2020} and \cite{aoyagi2021coexisting} research the relationship between information sets of traders, LP returns and the choice between centralized and decentralized exchanges. \cite{aigner2021} analyze the risk profile of LPs and discuss differences in the presence of concentrated liquidity. \cite{milionis2022automated} propose a decomposition of the LP returns into an instantaneous market risk component and a predictable component, which they term as "loss-versus-rebalancing". \cite{heimbach2021} analyze the returns of LPs of Uniswap V2 pools and study the movement of liquidity between pools. \cite{cartea2022decentralised} introduce the concept of predictable loss and apply it to CPMMs with concentrated liquidity. Additionally, \cite{priceAccuracyV3_2022} analyze the ability of Uniswap V3 to handle unexpected price shocks. 


\section{Framework}
\label{sec:framework}

In our paper we focus on a special case of AMMs: CPMMs with two assets. Let us denote the pool amounts of these two assets by $x$ and $y$. The reserve constraint is shown in equation {\eqref{eq:xyk}, where $k$ is the constant product of token amounts $x$ and $y$. There are three explicit actions that affect the pool's token reserves: \emph{Swap} one token for the other, provide liquidity in the form of $x$- and $y$-tokens to the pool (\emph{mint}) and redeem liquidity from the pool (\emph{burn}).

\begin{equation}
	x \cdot y = k
	\label{eq:xyk}
\end{equation}

\subsection{Token to Token Swap}
Assume that an agent wants to trade $x$-tokens for $y$-tokens. The agent sends $\Delta(x)$ of their $x$-tokens to the liquidity pool contract and receives $\Delta(y)$ $y$-tokens in exchange. The reserve constraint from \eqref{eq:xyk} must still hold for new token reserves after the trade.

We follow the notation of \cite{zcpFormal2018} and define $\alpha = \frac{\Delta(x)}{x}$ and $\beta = \frac{\Delta(y)}{y}$. Assume $\Delta(x)$ to be fixed, i.e., the trader provides a fixed amount of $x$-tokens to the pool. The token reserves after the trade $y'$ and change in token reserves $\Delta(y)$ are defined as:
\begin{align}
	y' = \frac{1}{1+\alpha}\cdot y \text{ ,} \hspace{2.5em}
\Delta(y) = \frac{\alpha}{1+\alpha}\cdot y
  	\label{eq:tokenreserveyaftertrade}
\end{align}

Note that the same relation applies for the opposite case where $\Delta(y)$ is fixed.

Relative prices are given by the first derivative of $k$ at a given point, allowing us to write $P_x=\frac{y}{x}$ and $P_y=\frac{x}{y}$ respectively. Considering the constant product constraint, only points on $k$ are feasible. This has two consequences. First, when someone uses the CPMM to swap $x$ for $y$-tokens, there is a diminishing marginal return for every $x$-token that is sent to the smart contract. Second, for any given amount of $x$-tokens, the contract will be able to quote a relative exchange rate and an amount in $y$-tokens that does not completely deplete its reserves.

\subsection{Liquidity Provision and Redemption}

Anyone can contribute liquidity to the pool by providing $n$ $x$-tokens and $n \cdot \frac{y}{x}$ $y$-tokens to the smart contract. This changes the token amounts as well as the constant $k$. The ratio between the two tokens remains unchanged. Both reserves are increased by the factor $\varphi = \frac{\Delta(x)}{x} = \frac{\Delta(y)}{y}$ as shown in \eqref{eq:liquidityprovision}.
\begin{align}
	k' = (1+\varphi)^2 k
  \label{eq:liquidityprovision}
\end{align}

In return, the LP receives (mints) a corresponding amount of liquidity tokens that represent partial pool ownership and can be redeemed for their proportional share of the pool's token holdings. Redemption is referred to as a burn action. It is the exact opposite of a \emph{mint} action and reduces the pool's $k$-value. 

\subsection{The Role of Trading Fees}
To incentivize liquidity provision, the model relies on trading fees. Let us assume a fee $\rho \in [0,1)$ with $\gamma := 1 - \rho$. The fee is charged on every trade and added to the liquidity pool and therefore leads to an increase in the normalized $k$, i.e., the $k$-value in relation to the outstanding liquidity tokens. 
Applying a fee to both equations in \eqref{eq:tokenreserveyaftertrade} leads to the following equations for the new token reserves as well as the change in token reserves:
\begin{align}
y'_\rho = \frac{1}{1+\alpha \gamma} \cdot y \text{ ,} \hspace{2.5em}    \Delta(y_\rho) = \frac{\alpha \gamma}{1+\alpha \gamma} \cdot y
  \label{eq:tokenreserveyaftertradewithfees}
\end{align}
%

Let us revisit the swap action and assume that an agent wants to trade $x$-tokens for $y$-tokens. They send $\Delta(x)$ of their $x$-tokens to the smart contract. In the setup with fees, the agent receives only $\Delta(y_\rho)$ $y$-tokens.

Since the smart contract adds the fee to the liquidity pool, the trade leads to an increase in $k$. 
These equations build the foundation for our LP profitability analysis.

\section{Formal LP Profitability Analysis}
\label{sec:profitabilityAnalysis}

To determine whether liquidity provision is profitable, we compare its return to the outcome of a pure hold strategy, where the investor maintains their initial allocation of $x$ and $y$. 
Formally, the buy and hold value in $x$-terms can be expressed as
\begin{align}
	x_0 + y_0 \frac{x_1}{y_1},
	\label{eq:BuyAndHold}
\end{align}

where the indices $0$ and $1$ represent the start and end points of the comparison period. Similarly, the liquidity pool investment value in $x$-terms can be expressed as
\begin{align}
	x_1 + y_1 \frac{x_1}{y_1} = 2 x_1.
	\label{eq:InvestLP}
\end{align}

The liquidity provision is worthwhile if \eqref{eq:InvestLP} $>$ \eqref{eq:BuyAndHold}. The difference between the two strategies arises from two distinct sources.

\emph{First}, liquidity provision has a non-negative fee accumulation effect. Whenever someone swaps assets using the CPMM, they will pay a trading fee. These fees are assigned proportionally to all LPs. 

\emph{Second}, any relative price shift away from the initial price ratio leads to a negative effect on the return compared to a pure hold strategy \cite{pintail2019}. Intuitively, this is a result of the passive market making and a quasi-arbitrage effect, i.e., the LP will own less of the more valuable and more of the less valuable token. This phenomenon is commonly referred to as \emph{divergence}- or \emph{impermanent loss}. 
The combination of these two effects leads to four possible cases. 

\subsection*{\textbf{Case 1:} $k_0 = k_1$, $\frac{x_0}{y_0} = \frac{x_1}{y_1}$}

It can be easily shown that  \eqref{eq:BuyAndHold} = \eqref{eq:InvestLP}.

\subsection*{\textbf{Case 2:} $k_0 = k_1$, $\frac{x_0}{y_0} \neq \frac{x_1}{y_1}$}
Case 2 represents a pure divergence loss $D$. Hence, it can be shown that \eqref{eq:BuyAndHold} $>$ \eqref{eq:InvestLP}. 
\begin{align}
	D &:= \frac{ x_1 + y_1 \frac{x_1}{y_1}}{ x_0 + y_0 \frac{x_1}{y_1}} - 1
	\label{eq:firstStepD}
\end{align}

We can now rewrite \eqref{eq:firstStepD} in $x_1$-terms and expand the fraction by $x_0$. 
%
%
%
Assuming $k_0 = k_1$, we get $x_0 \cdot y_0 \equiv x_1 \cdot y_1$, which can be rearranged to $\frac{x_1}{x_0} \equiv \frac{y_0}{y_1}$. We use this relation to expand our equation 
and take the square root.
\begin{align}
	D = 2 \cdot \frac{\sqrt{ \frac{x_1 \cdot y_0}{y_1 \cdot x_0}}}{1+ \frac{x_1 \cdot y_0}{y_1 \cdot x_0}}-1
\end{align}

Note that $\frac{x_1 \cdot y_0}{y_1 \cdot x_0}$ corresponds to the change in the price ratio between the two assets.

\subsection*{\textbf{Case 3:} $k_0 < k_1$, $\frac{x_0}{y_0} = \frac{x_1}{y_1}$}
It can be shown that \eqref{eq:BuyAndHold} $<$ \eqref{eq:InvestLP}. 
Starting from \eqref{eq:firstStepD}, rewriting the equation in $x_1$-terms, expanding it with $x_0$ and replacing $x_1$ with $x_0 \cdot \sqrt{\frac{k_1}{k_0}}$ and $y_1$ with $y_0 \cdot \sqrt{\frac{k_1}{k_0}}$.

\begin{figure}[]
	\begin{center}
		\input{profitabilityFrontierWithPriceLimits.tex}
	\end{center}
	\caption{Price limits of profitability frontier for a given $k_1$}
	\label{fig:profitabilityFrontierWithPriceLimits}
\end{figure}
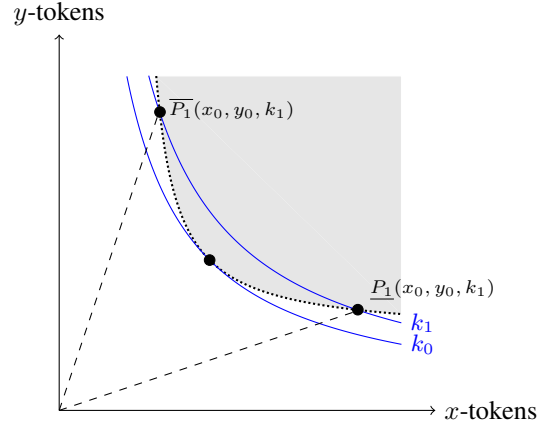

\subsection*{\textbf{Case 4:} $k_0 < k_1$, $\frac{x_0}{y_0} \neq \frac{x_1}{y_1}$}
%

Equating \eqref{eq:BuyAndHold} and \eqref{eq:InvestLP} and solving for $y_1$ we get
\begin{align}
	y_1 &= \frac{y_0 x_1}{2 x_1 - x_0}.
	\label{eq:y1ProfitabilityFrontier}
\end{align}

With both effects present, the profitability of the investment depends on which of the two effects is more pronounced. Equation \eqref{eq:y1ProfitabilityFrontier} can be interpreted as the profitability frontier with a convex profitability set. The function allows us to visualize the profitability frontier and the profitability set in $x$$y$-space, as shown in Figure \ref{fig:profitabilityFrontierWithPriceLimits}.

Figure \ref{fig:profitabilityFrontierWithPriceLimits} shows the profitability frontier (dotted curve) and the profitability space (shaded gray). For any given $k_1$ with $k_1 > k_0$, there is a set of price ratios (asset allocations) between $\overline{P_1}$ and $\underline{P_1}$ that represents profitable outcomes. 

From \eqref{eq:y1ProfitabilityFrontier} we can derive the limits of the profitability frontier.
	\begin{align}
		\lim\limits_{x_1 \rightarrow \infty}{\frac{y_0 x_1}{2 x_1 - x_0}} &= \frac{y_0}{2}
	\end{align}

\subsection{Effect of Mint Fees}
Transactions on Ethereum are subject to network fees as described by \cite{woodEthereum2014}. These network fees must not be confused with the trading fees described earlier. They are paid for every transaction on the Ethereum network. All three explicit actions that affect CPMM reserves (i.e., \emph{swap}, \emph{mint} and \emph{burn}) are transactions and therefore subject to a network fee.

Network fees are denoted in units of \emph{gas}. Every operation has a universally agreed upon cost in terms of gas units. When a transaction calls a smart contract function, it essentially executes all operations that are part of this function. Hence, each contract call has a gas cost associated with it.

\begin{figure}[]
	\begin{center}
		\input{profitabilityFrontierGasFees.tex}
	\end{center}
	\caption{Profitability frontier after mint fees}
	\label{fig:profitabilityFrontierWithGasFees}
\end{figure}
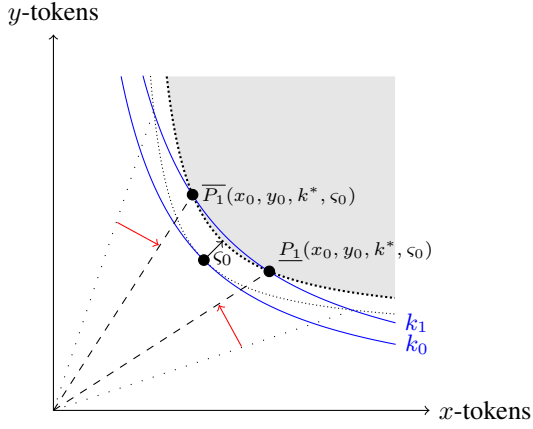

Transaction fees are computed by multiplying these gas units with a per unit price. The per unit price consists of a global base fee that adjusts dynamically to the demand in block space, as well as a voluntary tip, which is chosen by the transaction sender. A higher tip increases the probability that the transaction will get included in the next block and therefore decreases the expected confirmation time. In the context of decentralized exchanges, this is of particular importance as transactions are usually time critical.
Providing liquidity to a CPMM and closing the position requires multiple transactions. Including the ERC20 \emph{approve} calls this can take up to five blockchain transactions. The corresponding transaction fees can lower the profitability of the liquidity provision significantly and shift the profitability set. 


Let us denote the mint fee by $\varsigma_0$. Equating \eqref{eq:BuyAndHold} and \eqref{eq:InvestLP}, and incorporating mint fees, we obtain
\begin{align}
	\left(x_0 + y_0 \frac{x_1}{y_1}\right) \cdot \left(1 + \frac{\varsigma_0}{2x_0}\right) &= 2x_1. 
	\label{eq:profitabilityFrontierWithMintStart}
\end{align}	

From \eqref{eq:profitabilityFrontierWithMintStart} we can derive the shifted profitability frontier as a function of network fees
\begin{align}
	y_1 &= \frac{y_0 x_1 \cdot (1 + \frac{\varsigma_0}{2x_0})}{2x_1 - x_0 - \frac{\varsigma_0}{2}} 
	\label{eq:profitabilityFrontierWithMintSolution}
\end{align}

as well as the corresponding limits
\begin{align}
	\lim\limits_{x_1 \rightarrow \infty}{\frac{y_0 x_1 \cdot (1 + \frac{\varsigma_0}{2x_0})}{2x_1 - x_0 - \frac{\varsigma_0}{2}}} &= \frac{y_0 \cdot (1+\frac{\varsigma_0}{2x_0})}{2}.
	\label{eq:limitProfitabilityFrontierWithMint}
\end{align}

The effects on the profitability frontier are visualized in Figure \ref{fig:profitabilityFrontierWithGasFees}. Note that the shift leads to a narrower section of the $k$-indifference curve between $\overline{P_1}$ and $\underline{P_1}$. 

\begin{figure}[]
	\begin{center}
		\input{profitabilityFrontierBurnFees}
	\end{center}
	\caption{Profitability Frontier after burn Fees, with symmetric (blue) and asymmetric (red) burn fees.}
	\label{fig:profitabilityFrontierBurnFee}
\end{figure}
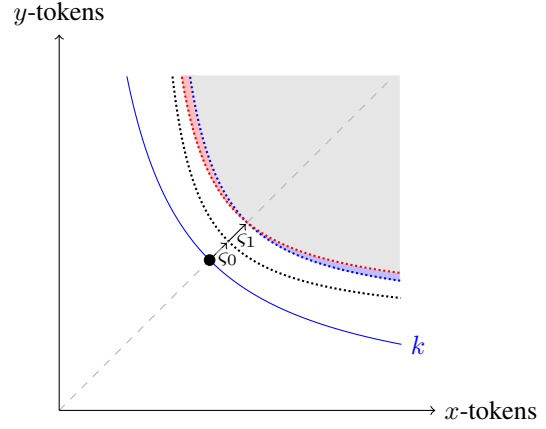

\subsection{Effect of Burn Fees}

Similarly to mint fees, LPs are subject to a burn fee. We denote this fee by $\varsigma_1$. It must be interpreted as an expected value. The exact fee depends on the base fee and the mempool competition at the time of the burn transaction. 
Moreover, effective fee size will be affected by the relative prices of the two pool tokens compared to the native protocol asset Ether (ETH). If the pool token prices increase (decrease) relative to ETH, the flat fee will have a smaller (larger) proportional effect. 

For CPMM pools without wrapped ETH (wETH), an ERC20 compliant version of ETH, we assume that the normalized pool value remains constant to ETH (symmetric case). The reasoning behind this assumption is that the prices at the time of minting are the best predictor for prices at a later point in time. As such, any profitability expectations should be based on these prices. 
For CPMM pools with wETH, we have additional information. Since one of the two tokens can be redeemed for ETH at a fixed 1:1 ratio, we know the prices of both pool tokens in ETH (asymmetric case). 
The two cases will yield slightly different profitability frontiers.
Let us first consider the symmetric case, where wETH is not part of the pool.
\begin{align}
	\left(x_0 + y_0 \frac{x_1}{y_1}\right) \cdot \left( 1+\frac{\varsigma_0 + \varsigma_1}{2x_0}\right) &=  2x_1.
	\label{eq:burnFeeStartK}
\end{align}
	Solving \eqref{eq:burnFeeStartK} for $y_1$ we get
\begin{align}
	y_1 &= \frac{y_0 x_1 \cdot (1 + \frac{\varsigma_0 + \varsigma_1}{2x_0})}{2 x_1 - x_0 - \frac{\varsigma_0 + \varsigma_1}{2} }
	\label{eq:profitabilityFrontierStartK}
\end{align}
and the corresponding limit
\begin{align}
	\lim\limits_{x_1 \rightarrow \infty}{ \frac{y_0 x_1 \cdot (1 + \frac{\varsigma_0 + \varsigma_1}{2x_0})}{2 x_1 - x_0 - \frac{\varsigma_0 + \varsigma_1}{2} }} &= \frac{y_0 \cdot (1+\frac{\varsigma_0 + \varsigma_1}{2x_0})}{2}.
	\label{eq:limitProfitabilityFrontierWithStartK}
\end{align}

For the asymmetric case, in which one of the pool tokens is wETH, the profitability frontier with $\varsigma_1 > 0$ will look slightly different. Let us assume that $y$ is wETH. It has a constant exchange rate to ETH, the asset in which the fees must be paid. Hence, the relative fee will change with the price ratio. We can derive this property from
\begin{align}
	\left(x_0 + y_0 \frac{x_1}{y_1}\right) \cdot \left( 1+\frac{\varsigma_0}{2x_0}\right) &=  2x_1 -\varsigma_1 \cdot \frac{x_1 y_0}{x_0 y_1}.
	\label{eq:burnFeeStartY}
\end{align}
Solving \eqref{eq:burnFeeStartY} for $y_1$ we get
\begin{align}
	y_1 &= \frac{y_0 x_1 \cdot \left(1 + \frac{\frac{\varsigma_0}{2} + \varsigma_1 }{x_0} \right)}{2 x_1 - x_0 - \frac{\varsigma_0}{2} }
	\label{eq:profitabilityFrontierStartY}
\end{align}
and the corresponding limits:
\begin{align}
	\lim\limits_{x_1 \rightarrow \infty}\frac{y_0 x_1 \cdot \left(1 + \frac{\frac{\varsigma_0}{2} + \varsigma_1 }{x_0} \right)}{2 x_1 - x_0 - \frac{\varsigma_0}{2} }={ \frac{y_0 \cdot \left(1 + \frac{\frac{\varsigma_0}{2} + \varsigma_1 }{x_0} \right) }{2} } 
	\label{eq:limitProfitabilityFrontierStartY}
\end{align}

\begin{align}
	\lim\limits_{y_1 \rightarrow \infty}\frac{y_1 \cdot (x_0+ \frac{\varsigma_0}{2})}{2y_1 - y_0 \cdot \left(1 + \frac{\frac{\varsigma_0}{2} + \varsigma_1 }{x_0} \right)}={\frac{x_0 + \frac{\varsigma_0}{2}}{2}}. 
	\label{eq:limitProfitabilityFrontierStartY}
\end{align}

Figure \ref{fig:profitabilityFrontierBurnFee} visualizes the shift of the profitability frontier for a given $\varsigma_1$. The blue profitability frontier represents the symmetric case, where wETH is not part of the pool. The red profitability frontier represents the asymmetric case, where wETH is part of the pool. The blue (red) shaded profitability set is exclusive to the symmetric (asymmetric) case.



%
\begin{figure}[]
	\center
	\includegraphics[width=0.95\columnwidth]{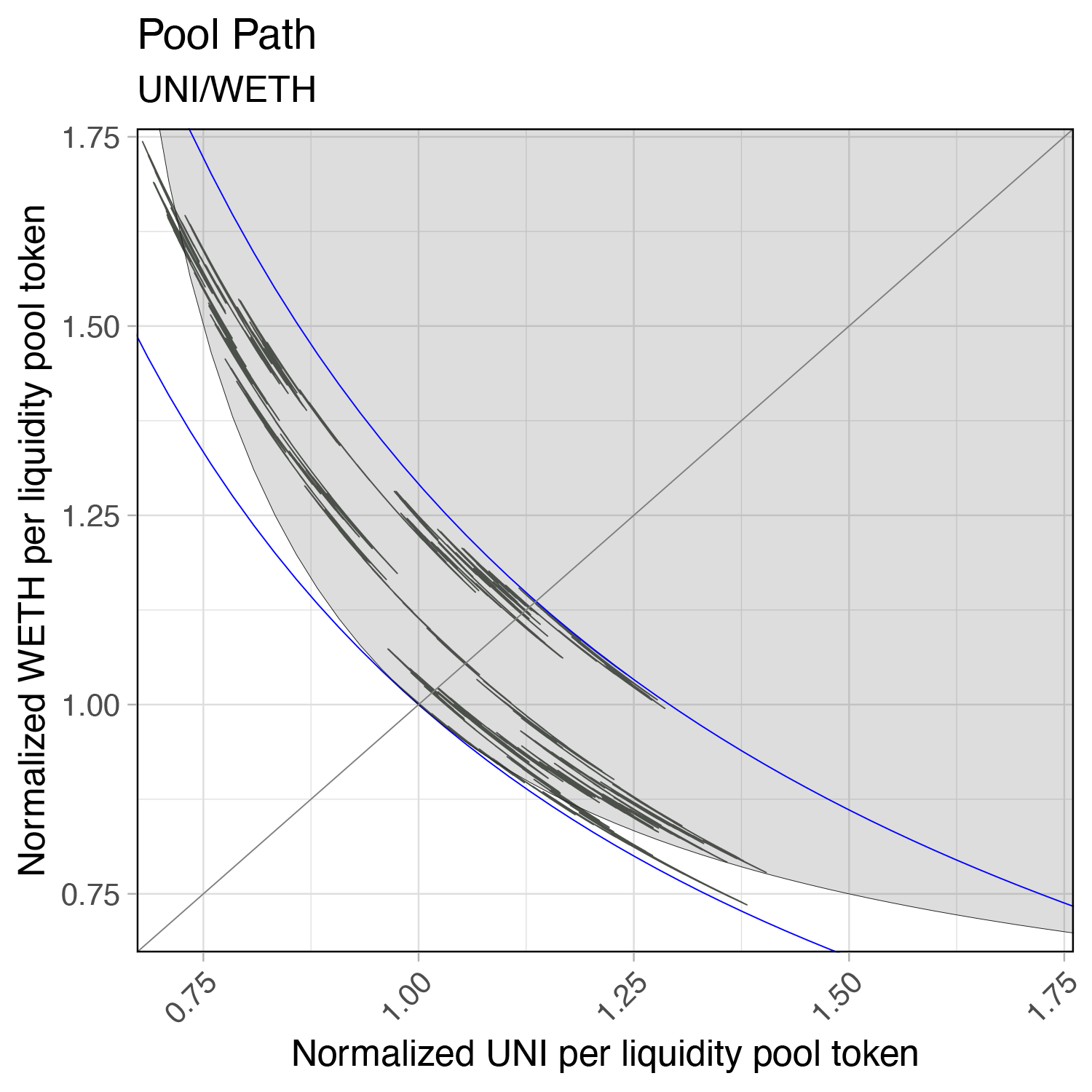}
	\caption{Pool path, block 11.073.311 to 13.390.744 (360 days).}
	\label{fig:poolPathUNIETH}
\end{figure}	
\begin{figure}[]
	\center
	\includegraphics[width=0.95\columnwidth]{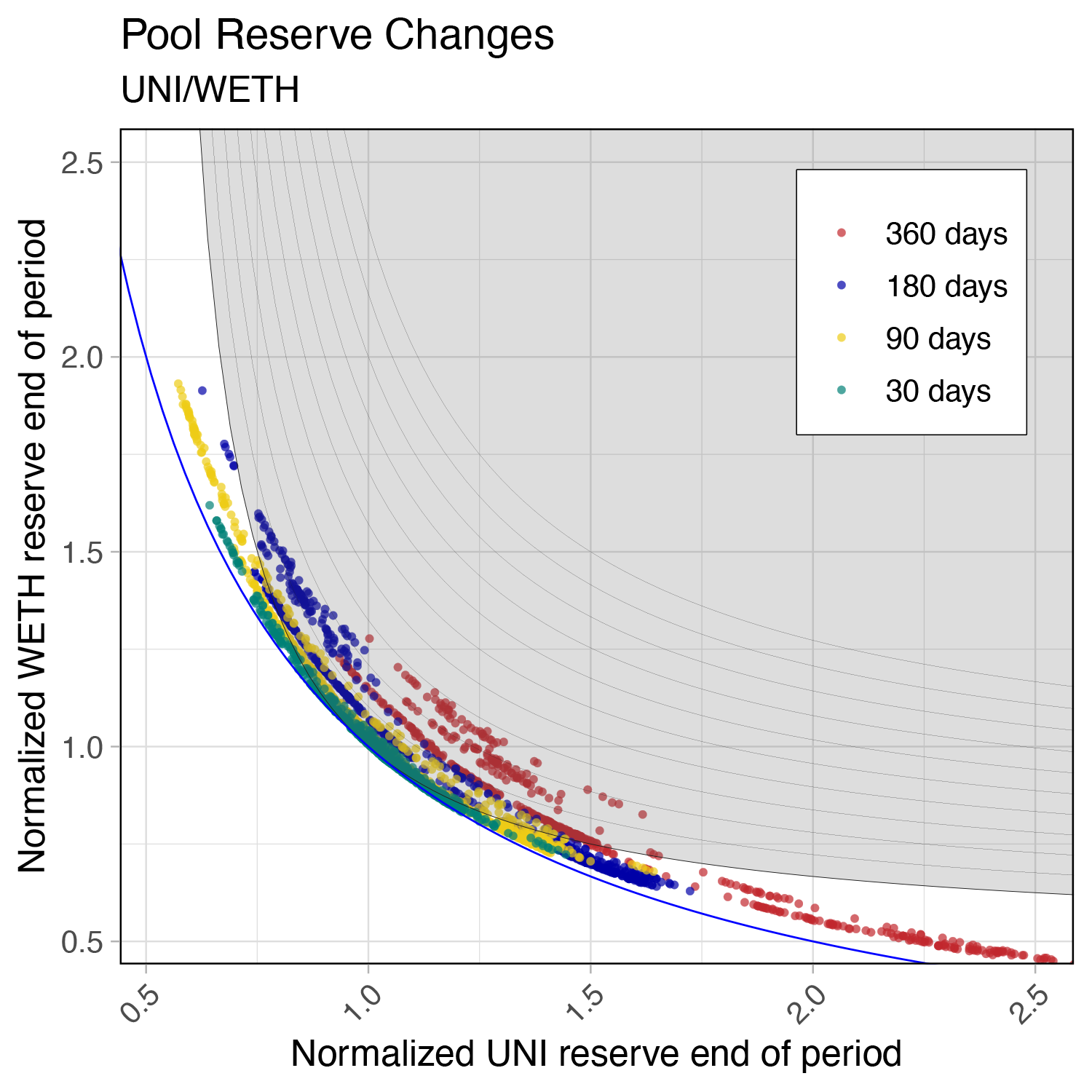}
	\caption{Reserve changes for different holding periods with relative combined fees of 0\% to 50\% of initial pool value (steps of 5\%). Fees paid in $y$-token.}
	\label{fig:reserveChangesUNI_WETH}
\end{figure}	
\begin{figure}[]
	\center
	\includegraphics[width=0.95\columnwidth]{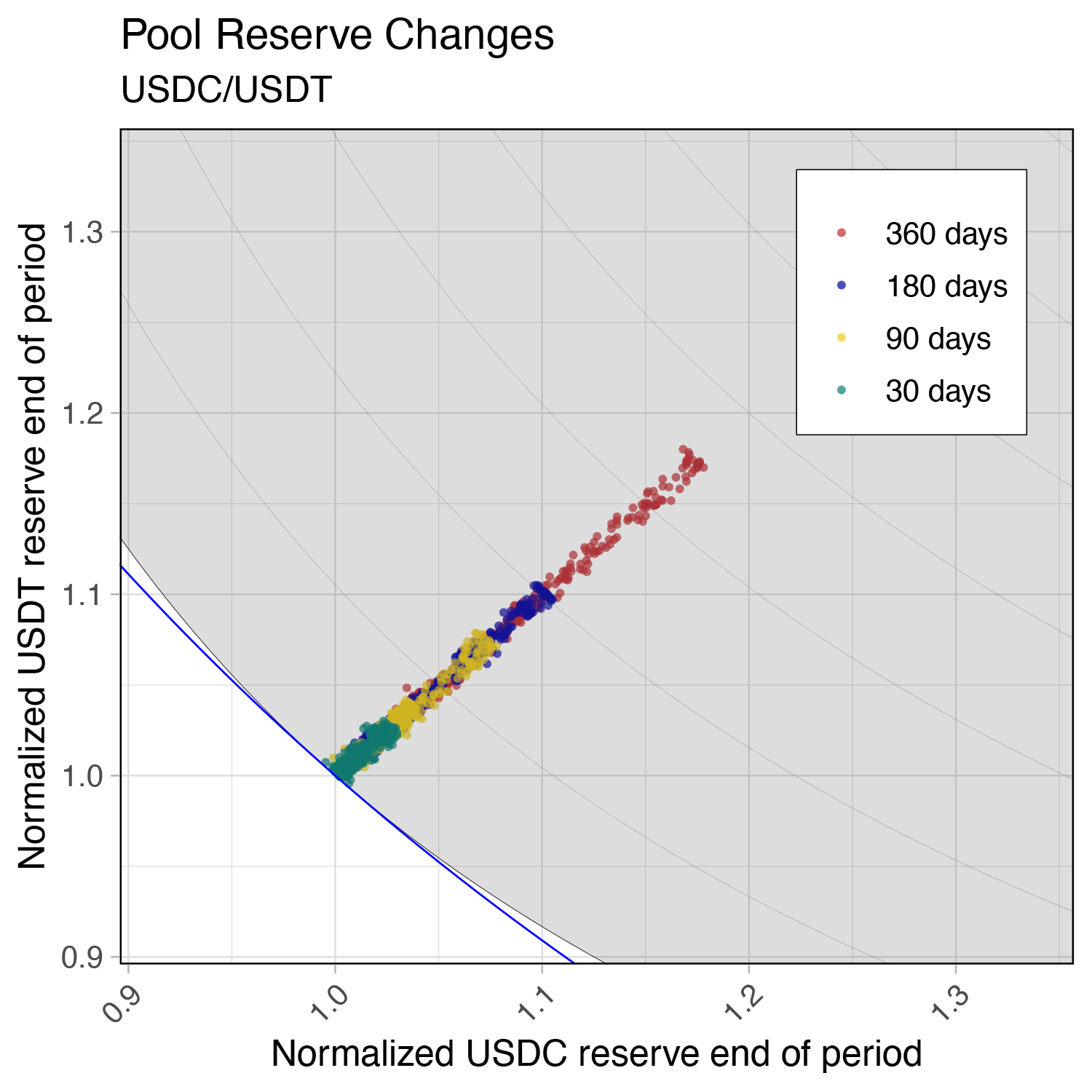}
	\caption{Reserve changes for different holding periods with relative combined fees of 0\% to 35\% of initial pool value (steps of 5\%). Fees constant in $k$.}
	\label{fig:reserveChangesUSDC_USDT}
\end{figure}	

\begin{table*}[t]
\vspace{0.1in}
\centering
\scalebox{0.95}{
\begin{tabular}[t]{ccccccccccccccc}
\hline \hline
		\multirow{2}{*}{\textbf{Pair}} & \multirow{2}{*}{\textbf{Fee}} & \multirow{2}{*}{\textbf{Type}} & \multirow{2}{*}{\textbf{N}} & \multicolumn{3}{c}{\textbf{30 Day Profitability}} && \multicolumn{3}{c}{\textbf{180 Day Profitability}} && \multicolumn{3}{c}{\textbf{360 Day Profitability}} \\ \cline{5-7} \cline{9-11} \cline{13-15}
		
		   &&&& Small & Med & Large && Small & Med & Large && Small & Med & Large \\ \cline{5-7} \cline{9-11} \cline{13-15}
&&&&&&&&&&&&&\\
		
USDC/WETH & Asymmetric & OpenMarket & 964 & 0.00 & 0.05 & 0.49 &  & 0.20 & 0.55 & 0.69 &  & 0.44 & 0.47 & 0.53 \\ 
DAI/WETH & Asymmetric & OpenMarket & 956 & 0.00 & 0.03 & 0.31 &  & 0.07 & 0.36 & 0.61 &  & 0.34 & 0.44 & 0.47 \\ 
WBTC/WETH & Asymmetric & OpenMarket & 951 & 0.00 & 0.00 & 0.12 &  & 0.00 & 0.19 & 0.64 &  & 0.00 & 0.09 & 0.72 \\ 
LINK/WETH & Asymmetric & OpenMarket & 951 & 0.00 & 0.04 & 0.34 &  & 0.09 & 0.23 & 0.78 &  & 0.17 & 0.38 & 0.54 \\ 
MKR/WETH & Asymmetric & OpenMarket & 951 & 0.00 & 0.00 & 0.19 &  & 0.06 & 0.28 & 0.77 &  & 0.23 & 0.51 & 0.85 \\ 
USDT/WETH & Asymmetric & OpenMarket & 951 & 0.00 & 0.06 & 0.52 &  & 0.32 & 0.61 & 0.70 &  & 0.47 & 0.52 & 0.60 \\ 
YFI/WETH & Asymmetric & OpenMarket & 890 & 0.02 & 0.05 & 0.31 &  & 0.10 & 0.35 & 0.77 &  & 0.10 & 0.35 & 0.47 \\ 
UNI/WETH & Asymmetric & OpenMarket & 829 & 0.00 & 0.00 & 0.10 &  & 0.01 & 0.14 & 0.49 &  & 0.11 & 0.42 & 0.61 \\ 
AAVE/WETH & Asymmetric & OpenMarket & 813 & 0.00 & 0.00 & 0.20 &  & 0.09 & 0.36 & 0.62 &  & 0.11 & 0.36 & 0.53 \\ 
DAI/USDC & Symmetric & Stable & 956 & 0.00 & 0.00 & 0.14 &  & 0.00 & 0.23 & 0.44 &  & 0.12 & 0.33 & 0.61 \\ 
USDC/USDT & Symmetric & Stable & 951 & 0.00 & 0.00 & 0.24 &  & 0.03 & 0.30 & 1.00 &  & 0.18 & 0.52 & 1.00 \\ 
DAI/USDT & Symmetric & Stable & 945 & 0.00 & 0.00 & 0.13 &  & 0.00 & 0.27 & 0.52 &  & 0.16 & 0.37 & 0.91 \\ 
\hline \hline \vspace{0.1em}
\end{tabular}
}
\caption{The table shows the percentage of observations that were profitable for the 12 largest pools (from the 11th day after pool deployment until end of 2022) by holding period and relative fee size (initial position).}
\label{table:toppooloverview}
\vspace{-0.8em}
\end{table*}%

\section{Empirical Analysis}
\label{sec:data}

In this section we use blockchain data to study the profitability frontier empirically. Our data set includes all Uniswap V2 transactions. We used the APIs of Infura and Alchemy to collect uniswap transactions, and the Coingecko API to gather the historical market capitalization of all listed tokens.
The data set includes observations from 132,657 pools, with a total of 2,371,811 \emph{mint}, 1,048,613 \emph{burn} and 93,749,446 \emph{swap} events for the period between block number 10,000,835 (4th of May 2020, deployment of the Uniswap V2 factory contract) and block number 16,308,189 (last block of 2022).  
To understand the data it may be useful to explore an example. Figure \ref{fig:poolPathUNIETH} shows one \emph{mint} event (initial point) and the corresponding pool path in $xyk$-space for the subsequent 360 days, normalized w.r.t. the initial allocation, i.e., the starting point is $(1, 1)$.

To generalize the analysis, we assume different holding periods and compute outcomes of virtual liquidity positions. We take daily pool reserve snapshots (noon UTC) and divide them by the outstanding amount of liquidity pool tokens. We then observe for each day the reserve per liquidity pool token after different holding periods (30, 90, ..) and set this in relation to the initial reserves.

Figure \ref{fig:reserveChangesUNI_WETH} shows the outcome of each observation for the UNI/WETH ``open market'' pool. Again, the starting point of each observation is $(1, 1)$ and each point reflects the relative outcome of one observation. The profitability space without gas fees is highlighted as the gray area and the gray lines show the profitability frontier for different assumptions of relative network fees. Note, that fees are paid in wETH, which is part of the pool. Hence, the profitability frontiers are visualizations of \eqref{eq:profitabilityFrontierStartY}. Points outside the gray area are not profitable even if relative gas fees were zero. The profitability of points that lay within the gray area depends on the relative fee value. 
The various profitability frontiers depict relative fees in steps of 5 percentage points of the initial allocation.

Figure \ref{fig:reserveChangesUSDC_USDT} shows an example pool with two stablecoins, which are expected to trade within a narrow price range. The results demonstrate that, in the absence of network fees, a majority of observations are profitable. However, when substantial relative network fees are present, short-term liquidity provisions may not be profitable.

A selection of pools were chosen for Table \ref{table:toppooloverview} based on the following criteria: (1) The pool has at least 100,000 events (\emph{mint}, \emph{burn} and \emph{swap} combined), and (2) both tokens of the pool maintained a relevant market capitalization for at least 24 months. We took a snapshot of the market capitalizations on the 15th day of each month (as recorded on Coingecko) and compared it to the market capitalization of ETH. A token was considered relevant if it maintained at least 0.5\% of the Ether market capitalization. 

Table \ref{table:toppooloverview} summarizes the profitability of liquidity provision for these pools. The table shows the percentage of observations that were profitable for different holding periods (30, 180, and 360 days) and liquidity position sizes (small, medium, and large). The liquidity position sizes are defined as the ratio of fees to the position size, with small (medium, large) representing an implied fee share of 0.10 (0.05, 0.01) of the initial holdings, with equally divided fees for \emph{mint} and \emph{burn} events.
The results indicate that the profitability of liquidity provision in CPMM pools is heavily dependent on the size of the position and the holding period. Small and medium-sized positions are generally not profitable in the short-term, while larger positions have a higher likelihood of profitability. Additionally, the volatility of the underlying assets plays a significant role, as higher volatility in ``open market'' pairs can provide more opportunities to offset the high relative gas fees, but the higher volatility also comes with higher divergence loss risk.
It would be of interest to include a category for ``Pure Trend'' pools, however, the authors have not been able to identify a pool with significant liquidity that fits this category. Such pools would likely exhibit characteristics similar to ``Stable'' pairs, with the added factor of a consistent trend in the price ratio, creating a predictable, trend-based divergence loss.

\section{Conclusion}
\label{sec:conclusion}
In this paper we analyzed LP profitability in CPMMs. In the first part we derived the profitability frontier and the corresponding profitability set and formalized the effect of \emph{mint} and \emph{burn} fees. 
Based on these findings, we show that small liquidity positions face an implicit lock-in. This effect becomes more pronounced for higher relative network fees. 
In the second part of the paper we analyzed a large data set and studied the profitability frontier empirically. Empirical evidence backs the expected results and allows us to visualize pool profitability in $xyk$-space. 

The paper provides a novel analytical framework to study LP profitability and highlights the importance of layer 2 deployments and improvements to the token approval process. These measures will decrease relative transaction fees and thereby reduce holdup problems for small-scale LPs.

\section*{Acknowledgments}

The authors would like to thank Felix Bekemeier, Florian Bitterli, Dario Thürkauf, Mitchell Goldberg, Emma Middleton, Matthias Nadler, Remo Nyffenegger and Katrin Schuler
for their valuable inputs.

\bibliographystyle{IEEEtran} 
\bibliography{mybib.bib}

\end{document}

%% file: ProfitabilityFrontierWithPriceLimits.tex
\begin{tikzpicture}[]
 
	\draw[color = white, fill = gray!20] (1.29,4.45) -- (4.55,4.45) -- (4.55,1.28) ;
   
	\draw[samples = 200, scale=1, xshift = 0cm, yshift = 0cm, domain=1.29:4.55, thick, densely dotted, smooth, variable=\x, black, fill = gray!20] plot ({\x}, {2*\x/(2*\x-2)});
    
	\draw[samples = 200, color=blue, scale=1, xshift = 0cm, yshift = 0cm, domain=0.9:4.55, smooth, variable=\x] plot ({\x}, {4/\x}) node[right,color=blue] {\small{$k_0$}} ;    

	\draw[samples = 200, color=blue, scale=1, xshift = 0cm, yshift = 0cm, domain=1.19:4.55, smooth, variable=\x] plot ({\x}, {5.3/\x}) node[right,color=blue] {\small{$k_1$}} ;    

	\draw[fill=black] (1.34,3.97) circle(2pt);
	\draw[fill=black] (3.97,1.34) circle(2pt);
  
	\draw[dashed] (0,0)--(1.34,4.01) node[right]{\scriptsize$\overline{P_{1}}(x_0,y_0,k_1)$} ;  
	\draw[dashed] (0,0)--(4.01,1.34) node[above right]{\scriptsize$\underline{P_{1}}(x_0,y_0,k_1)$} ;    
	
	\draw[->] (0,0)--(5,0) node[below,midway]{} node[right] {$x$-tokens} ;
	\draw[->] (0,0)--(0,5) node[above,midway,rotate=90]{} node[above] {$y$-tokens};
	
	\draw[fill=black] (2,2) circle(2pt);
	
\end{tikzpicture}

%% file: profitabilityFrontierGasFees.tex
 \begin{tikzpicture}[]
 

	\draw[samples = 200, scale=1, xshift = 0cm, yshift = 0cm, domain=1.29:4.55, densely dotted, smooth, variable=\x, black] plot ({\x}, {2*\x/(2*\x-2)});
   

	    \draw[color = white, fill = gray!20] (1.48,4.45) -- (4.55,4.45) -- (4.55,1.48) ;
	\draw[samples = 200, scale=1, domain=1.5065:4.54, thick, densely dotted, smooth, variable=\x, black, fill = gray!20] plot ({\x}, {(2*\x * (1 + (0.5/4)) )/ (2*\x - 2 - (0.5/2))});

	\draw[samples = 200, color=blue, scale=1, xshift = 0cm, yshift = 0cm, domain=0.9:4.55, smooth, variable=\x] plot ({\x}, {4/\x}) node[right,color=blue] {\small{$k_0$}} ;    

	\draw[samples = 200, color=blue, scale=1, xshift = 0cm, yshift = 0cm, domain=1.19:4.55, smooth, variable=\x] plot ({\x}, {5.3/\x}) node[right,color=blue] {\small{{$k_1$}}};    

	\draw[fill=black] (1.85,2.87) circle(2pt);
	\draw[fill=black] (2.87,1.85) circle(2pt);
  
	\draw[dashed] (0,0)--(1.85,2.87) node[right]{\scriptsize$\overline{P_{1}}(x_0,y_0,k^*,\varsigma_0)$} ;  
	\draw[dashed] (0,0)--(2.87,1.85) node[above right]{\scriptsize$\underline{P_{1}}(x_0,y_0,k^*,\varsigma_0)$} ;    
	
	\draw[loosely dotted] (0,0)--(1.34,4.01) node[right]{} ;  
	\draw[loosely dotted] (0,0)--(4.01,1.34) node[above right]{} ;   
	
	\draw[->] (0,0)--(5,0) node[below,midway]{} node[right] {$x$-tokens} ;
	\draw[->] (0,0)--(0,5) node[above,midway,rotate=90]{} node[above] {$y$-tokens};
	
	\draw[fill=black] (2,2) circle(2pt);

	\draw[->, black] (2,2)--(2.25,2.25) node[midway,xshift = 0.13cm, yshift = -0.1cm,black]{\small{$\varsigma_0$}};

	\draw[->, red] (2.5,0.85)--(2.2,1.4);
	\draw[->, red] (0.85,2.5)--(1.4,2.2);

	
\end{tikzpicture}

%% file: profitabilityFrontierBurnFees.tex
 \begin{tikzpicture}[]

    


\filldraw[
          draw=red!25,
          fill=red!25,
          samples=50,
        ]
            plot[domain=1.635:2.5]
                ({\x}, { (\x * 2 * (1 + ((1)/(2) * ((0.5)/(2) + 0.5 ) )  ) / ( 2 *\x - 2 - ((0.5)/(2))) })
            -- cycle
        ;
       
\draw[color = red!25, fill = red!25] (1.62,4.45) -- (1.8,4.45) -- (1.8,4);

        \filldraw[
          draw=blue!25,
          fill=blue!25,
          samples=50,
        ]
            plot[domain=2.5:4.47]
               ({\x}, {(2*\x * (1 + (1/4)) )/ (2*\x - 2 - (1/2))})

            -- cycle
        ;

\draw[color = blue!25, fill = blue!25] (3.62,1.9) -- (4.52,1.9) -- (4.52,1.72);

\filldraw[
          draw=gray!20,
          fill=gray!20,
          samples=50,
        ]
            plot[domain=2.5:4.5]
                ({\x}, { (\x * 2 * (1 + ((1)/(2) * ((0.5)/(2) + 0.5 ) )  ) / ( 2 *\x - 2 - ((0.5)/(2))) })
            -- 
            plot[domain=1.74:2.5]
	            ({\x}, {(2*\x * (1 + (1/4)) )/ (2*\x - 2 - (1/2))})
            -- cycle
        ;

\draw[color = gray!20, fill = gray!20] (1.72,4.45) -- (4.52,4.45) -- (4.52,1.82);

    
    
        
	\draw[samples = 200, scale=1, domain=1.5065:4.54, thick, densely dotted, smooth, variable=\x, black] plot ({\x}, {(2*\x * (1 + (0.5/4)) )/ (2*\x - 2 - (0.5/2))});

    
	\draw[samples = 200, scale=1, domain=1.74:4.54, thick, densely dotted, smooth, variable=\x, blue] plot ({\x}, {(2*\x * (1 + (1/4)) )/ (2*\x - 2 - (1/2))});

    
	\draw[samples = 200, scale=1, domain=1.63:4.54, thick, densely dotted, smooth, variable=\x, red] plot ({\x}, { (\x * 2 * (1 + ((1)/(2) * ((0.5)/(2) + 0.5 ) )  ) / ( 2 *\x - 2 - ((0.5)/(2))) });    


      \draw[samples = 200, color=blue, scale=1, xshift = 0cm, yshift = 0cm, domain=0.9:4.55, smooth, variable=\x] plot ({\x}, {4/\x}) node[right,color=blue] {$k$} ;    


	\draw[->, black] (2,2)--(2.23,2.23) node[midway,xshift = 0.13cm, yshift = -0.1cm,black]{\small{$\varsigma_0$}};
	
	\draw[->, black] (2.25,2.25)--(2.48,2.48) node[midway,xshift = 0.13cm, yshift = -0.1cm,black]{\small{$\varsigma_1$}};

	\draw[black, dashed, black!30] (0,0)--(2.25,2.25);
	\draw[black, dashed, black!30] (2.5,2.5)--(4.4,4.4);        

  \draw[->] (0,0)--(5,0) node[below,midway]{} node[right] {$x$-tokens} ;
  \draw[->] (0,0)--(0,5) node[above,midway,rotate=90]{} node[above] {$y$-tokens};
  
  \draw[color=blue] (0.5,4.5) node[rotate= -75,above right]{};
  \draw[fill=black] (2,2) circle(2pt);
 \end{tikzpicture}